\input /usr/local/lib/tex/inputs/psfig.tex
\documentstyle[preprint,revtex]{aps}
\newcommand{\eqn}[2]
{\begin{equation}\label{#1} #2 \end{equation} }
\newcommand{\eno}[1]{Eq.~(\ref{#1})}
\newcommand{\ens}[2]{Eqs.~(\ref{#1}) and (\ref{#2})}
\newcommand{\gvec}[1]{\mbox{\boldmath $#1$}}
\newcommand{\uv}[1]{\widehat{{\bf #1}}}
\newcommand{\vecb}[1]{{\bf #1}}
\newcommand{\del}[0]{\partial}
\newcommand{\avg}[1]{\left\langle #1 \right\rangle}

\newcommand{\vr}[0]{{\bf r} }
\newcommand{\vk}[0]{{\bf k}}
\newcommand{\grad}[0]{\gvec{\nabla}}
\newcommand{\pp}[2]{\frac{\partial #1}{\partial #2} }
\newcommand{\dd}[2]{\frac{d #1}{d #2} }

\newcommand{\vB}[0]{{\bf B} }
\newcommand{\vE}[0]{{\bf E} }
\newcommand{\vA}[0]{{\bf A} }
\newcommand{\vu}[0]{{\bf u} }
\newcommand{\vje}[0]{{\bf J_e} }

\begin{document}
\pagestyle{empty}
\vspace{1cm}
\begin{title}
The Nonequilibrium Dynamics of Driven Line Liquids
\end{title}
\vspace{1cm}
\author{Terence Hwa}
\begin{instit}
Department of Physics\\
Harvard University\\
Cambridge, Massachusetts  02138
\end{instit}
\vfill
\begin{abstract}
We study the nonequilibrium dynamics of line liquids as realized in
the nonlinear motion of flux lines of a superconductor
 driven by an applied electric current. Our analysis
suggests  a transition in the dynamics of the lines
 from a smooth, laminar phase
at small driving forces, to a rough, turbulent phase when the
drive is increased. We explore the nature of these phases and
 describe interesting analogies to driven
diffusion and growing interfaces.
\end{abstract}
\vfill
\vskip 12pt
\newpage
\pagestyle{plain}
\pagenumbering{arabic}
\begin{narrowtext}

The statistical mechanics and dynamics of extended objects such as
 lines have
attracted increasing attention, especially in the context of
  flux lines of a superconductor\cite{ns,vg,nl,mn}, in recent years.
In this paper, we study the {\it nonequilibrium}
dynamics of a driven line liquid.
Our analysis suggests that such a
system can exhibit a transition from a ``laminar''
phase described by equilibrium fluctuations
at small driving forces to a ``turbulent'' phase dominated by
nonlinear effects as the driving force is increased.
We explore the nature of these phases and describe
interesting analogies to driven diffusion\cite{js}
 and growing interfaces\cite{kpz}.

Although the subject of our investigation is the driven dynamics
of a generic collection of lines in the liquid phase,
we shall develop our analysis in the context of the flux lines of a
superconductor for the sake of concreteness.
As is well known for type-II superconductors\cite{SC},
magnetic flux lines can penetrate a sample under the
application of an external magnetic field {\bf H}, forming an Abrikosov
lattice at low temperatures. However, as pointed out by
Nelson\cite{ns},
thermal fluctuations can destroy the translational order of the flux
lines, melting the flux lattice into a line liquid.
The equilibrium properties of the line liquid has been
studied in details by Nelson {\it et al}\cite{ns,nl,mn}.
This paper concerns the {\it nonequilibrium} dynamics of the
flux line liquid driven by an applied electric current $\vje\perp{\bf
H}$. But before delving into the driven dynamics, we shall
first give a brief review of the dynamics of the lines in
equilibrium.

We describe the long-wavelength, long-time behavior of the line
liquid by a vector field $\vB(\vr,t)$, with the
constraint $\grad\cdot\vB=0$. In the specific context of
superconductors, \vB has the natural interpretation of the
coarse-grained magnetic
field carried by the flux lines. However, it is worth noting that
the divergentless \vB-field lends itself naturally to the
description of a generic collection of lines without free ends.
Let the external field {\bf H} be in the $z$-direction, then the
$z$-component of \vB describes the  density of the flux lines
and the $x,y$- (or collectively called the $\times$-) components
describe the local tilt.
The dissipative thermal motion of the flux lines in the liquid phase
 destroys the superconductivity of the sample\cite{note0}.
The resulting dynamics
is diffusive and can be described by the Langevin equation\cite{mn},
$\del_t\vB = \nu \nabla^2\vB + \grad\times\gvec{\eta}$,
where we have scaled out the trivial anisotropy factors between
the $z$- and $\times$- directions. The Langevin noise is employed to
summarize thermal fluctuations over the microscopic degrees of
freedom.  It must take on the form of a curl to keep \vB
divergentless,
but is otherwise taken to be uncorrelated and Gaussian
distributed, with the second moment
$\avg{ \eta_i(\vr,t)\eta_j(0,0) } = 2D\delta_{ij}
\delta^3(\vr)\delta(t)$.
The choice of the noise spectrum is justified by matching
the resulting static structure factor,
$\avg{\vB(\vk,t)\vB(\vk',t)}$, with the
analogous quantity obtained from the equilibrium theory\cite{nl}.
The phenomenological parameters $D$ and $\nu$
can be identified with the temperature and elastic moduli of the
equilibrium theory.

We now consider nonequilibrium effects introduced by an
applied external current $\vje \perp \uv{z}$, taken here to be in the
$y$-direction. The external current provides a
Lorentz force $\vecb{F} = \vje\times\vB$ which drives the line
liquid in the $x$-direction. The viscous motion\cite{mn} of the
flux lines in turn
induces an electric field \vE which must be included in the equation
of motion. The most general form allowed by symmetry is
\eqn{4}{\vE = \rho(B)\vje + \tilde\rho(B)(\uv{B}\cdot\vje)\uv{B},}
where  $B$ and
$\uv{B}$ denote the norm and direction of \vB respectively, and
 the coefficients $\rho$,
$\tilde\rho$ can be identified as the field-dependent
resistivity\cite{note1}.

For simplicity, we make a reasonable assumption that
 the fluctuation of \vB in the direction of $\vje$ is small (i.e.,
$B_y \ll B_x,B_z$), and only consider the effect of the first
term in \eno{4}. We will later show the self-consistency of this
assumption.  The dynamics is then more conveniently expressed in
terms of the vector potential \vA. With $\grad\times\vA = \vB$,
we have
$$ \pp{\vA}{t} = \nu\nabla^2\vA  - \rho(\vB)\vje + \gvec{\eta} +
\grad\phi,$$ where the last term is fixed by the choice of gauge
$A_z=0$.
We look for the fluctuation of $\vA$ about the average by using a
{\it displacement} field, $\vu = \uv{z}\times(\vA - B_0 x\uv{y})$,
 where $B_0$ is the mean density of the lines.
Expanding $\rho(B)$ and keeping the
leading order nonlinearity, we obtain
\widetext
\eqn{5}{\pp{\vu}{t} = \nu\nabla^2\vu + \uv{T}\left[E_0 - v
(\grad_\times\cdot \vu)  +  \frac{\lambda_\times}{2}
(\grad_\times \cdot \vu)^2 +
\frac{\lambda_z}{2} \left(\pp{\vu}{z}\right)^2\right] + \gvec{\xi}.}

\narrowtext
In \eno{5}, $\uv{T} = \uv{J_e} \times \uv{z}$ is the
transport direction, $-\grad_\times \cdot \vu = B_z-B_0$ is the
density fluctuation, and $\del_z\vu = \vB_\times$ describes the
local tilt (see Ref. \cite{nl} and \cite{mn}).
The parameters in $[...]$ are given by the Taylor
expansion of $\rho(B)$ about $B_0$. For example,
 $E_0 = \rho(B_0)J_e$ is the mean electric field induced; it
produces an overall translation in $u_\|\equiv\vu\cdot\uv{T}$ and
can be shifted away. Similarly,
$v = \rho'(B_0)J_e$ is the mean drift rate of the lines; it is again
shifted away by a Lorentz boost in the $\uv{T}$ direction.
The nonlinear terms remaining are characterized by the coefficients
 $\lambda_\times =\rho''(B_0)J_e$
and $\lambda_z = \rho'(B_0)J_e/B_0$.
For superconductors, $\rho(B)$ is usually a
monotonically increasing function, saturating at large
values  of $B$ \cite{Iye}. This implies that $E_0, v$, and $\lambda_z$
are  always positive, while $\lambda_\times$ may take on either signs.
While the response of the system to a sign change in
$\lambda_\times$ is
an interesting study in itself\cite{wolf}, we shall be concerned
with the situation $\lambda_\times>0$ in this study.  Anticipating
the restoration of isotropy between the $z$- and $\|$- directions,
we set $\lambda_\times=\lambda_z=\lambda$ from here on.
Finally, the new noise is
$\gvec{\xi} = \uv{z}\times(\gvec{\eta} + \grad_\times\phi)$,
with $\del_z\phi = \eta_z$ from the gauge choice.
This gives $\avg{\xi_i(\vk,t)\xi_j(\vk',t')} =
2D_{ij}(\vk)\delta^3(\vk+\vk')\delta(t-t')$,
where the noise spectrum is
\eqn{5.2}{D_{ij}(\vk) = D\cdot\left[ \delta_{ij} + (k_\times^2
\delta_{ij} - k_\times^i k_\times^j)/k_z^2\right].}
It is worth
emphasizing that although \eno{5} is derived through some rather
specific considerations such as the form of $\rho(B)$, the
general form of the dynamics is purely a consequence of
the applied driving force
$\vje$ which breaks spatial isotropy. As a result, we expect \eno{5}
to be a good description for a generic line liquid with dynamics
local in \vB and \vu.

A  systematic investigation of the scaling behaviors is the
method of dynamic renormalization-group (DRG)\cite{DRG}.
In that approach, we generalize the space of
$\times$-directions (and thereby the components of the
displacement field \vu) from $2$- to $d_\times$- dimensions. [Among
the $d_\times$ directions, the drift direction $\uv{T}$ is special and
is called the $\|$-direction. The remaining directions are
collectively called the $\perp$-direction, with $d_\perp =
d_\times - 1$.]
Also we  generalize the $z$-direction from $1$- to
$d_z$- dimensions.
\eno{5} is studied near the critical dimension
$d = d_z + d_\times = 2$, and the result is then analytically
continue to the physical
dimension $(d_\times,d_z) = (2,1)$.
The most important output of the DRG analysis is a recursion
relation which describes the relevancy of the nonlinearity at
successively larger length scales.
In a one-loop study, the recursion relation has the form
\eqn{rg}{\dd{g}{l} = (2-d) g + I(d_z,d_\times) g^2,}
where the coupling constant $g \sim \lambda^2D/\nu^3$ characterizes
the effective strength of nonlinearity,
$I(d_z,d_\times)$ is the one-loop result evaluated
{\it along} the line of critical dimension $d_z+d_\times=2$, and
$l \sim \log(1/k) \to \infty$ is the infra-red limit of interest.

In the physical dimension $d=3$, the first term in \eno{rg} is
always negative, indicating the irrelevancy of a small nonlinearity.
However, the nonlinearity can become relevant if higher order terms
in the recursion relation are positive. For example, if $I(d=2)>0$,
then a dynamic phase transition occurs at $g_c
= (d-2)/I$. For $g<g_c$, the nonlinearity is still irrelevant. But
for $g>g_c$, the coupling constant flows to large values and the
system is described by a {\it strong coupling} fixed point\cite{dpt}.
Thus the behavior of the system at the physical dimension is
dictated to a large extent by  what happens right
at the line of critical dimensions, i.e., by the sign of $I(d=2)$.
To better understand the behavior of the system then,
we shall now consider some limiting cases on the critical line.

We first examine the case
with $(d_\times,d_z)=(2,0)$, corresponding to the situation where
an external field {\bf H} is applied normal to a thin-film
superconductor and the external current is applied in the plane.
 In this case, the flux lines
become point vortices and the problem becomes quite simple.
Since $\partial_z \vu = 0$, the configuration of the vortices can
be described by a scalar field $n= -\grad_\times\cdot\vu$ which is the
fluctuation of the vortex density. \ens{5}{5.2}
 then reduce to
the well-known driven-diffusion system (DDS)\cite{js}.
DRG method has been used to investigate the
scaling properties of that system. One finds $I(d_z=0, d_\times=2) <
0$, indicating the marginal irrelevancy of the nonlinearity.

We next investigate
the limit $(d_\times,d_z)=(1,1)$, or $d_\perp = 0$.
This corresponds to applying an
external field {\bf H} in the plane of a thin film
superconductor, with an applied electric current normal to the
plane as shown in Figure 1a. In this special configuration,
the displacement field $\vu$ is reduced to a scalar, $u_\|$.
 The noise
spectrum also simplifies as the nonlocal term vanishes.
\ens{5}{5.2}  then become the simpler anisotropic
Kardar-Parisi-Zhang (KPZ) equation which describes the growth of a
tilted crystalline surface\cite{kpz,wolf}.
The connection between the lines and interfaces is intuitively simple:
The lines shown in Figure 1a can be viewed as the contour plot
of a tilted surface\cite{duality}.
And the growth of such a surface corresponds
to the movement of the lines to the right. Mathematically, $u_\|$
describes the height fluctuation of an interface.

For the KPZ equation, one finds the one-loop term in \eno{rg} to be
positive\cite{kpz}.  So the nonlinearity is marginally
{\it relevant}, and the the asymptotic scaling
behavior is described by a strong coupling fixed point.
Extensive numerical studies\cite{surface}
 found that  fluctuations in $u_\|$
diverge algebraically with system size in the strong coupling phase.
In the interface language, this corresponds to the divergence of
interfacial width, and the resulting surface is called ``rough".
A contour plot of a rough surface yields a ``scrambled'' line
configuration with the proliferation of vortex loops (see Figure
1b).  In contrast, a smooth
interface give a set of roughly parallel lines without loops (Figure
1a).

The physical problem of a line liquid in 3-dimensions lies
somewhere in between the KPZ and the DDS limits.
To find out whether this problem may also flow to strong coupling,
we need to know the boundary between the KPZ- and DDS-
dominated regions in the space of generalized dimensions $(d_\times,
d_z)$. We can get an estimate by computing the function $I$ in
\eno{rg} and see where it changes its sign. However, a full DRG
treatment of \eno{5} is rather cumbersome.
Here we discuss a truncated version which
contains the essential physics.

In \eno{5}, we note that while $u_\|$ suffers renormalization by the
nonlinearities, $\vu_\perp$ is always described by the linear
theory. If the system flows to strong coupling, then only $u_\|$ will
pick up anomalous scaling and dominate the dynamics.  We can
therefore ignore terms like $\grad_\perp\cdot\vu_\perp$ and $\del_z
\vu_\perp$ in the dynamics of $u_\|$. (This is also the
justification for neglecting the cross term in \eno{4}.)
We also find the nonlocal part of the noise spectrum in
\eno{5.2} not to renormalize, while the local part of
$\avg{\xi_\| \xi_\|}$  does. So the nonlocal part can again be
ignored in the strong coupling limit. Consequently, the isotropy
between the $\|$- and $z$- directions is restored. To compactify
notations, we denote the
union of the two directions to be the $\dagger$-direction, and obtain
 the following simpler equation of motion,
\eqn{akpz}{ \pp{u_\|}{t} = \nu_{\dagger}\grad_{\dagger}^2 u_\| + \nu_\perp
\grad_\perp^2 u_\| + \frac{\lambda}{2} (\grad_{\dagger} u_\|)^2 + \xi_\|,}
with a constant noise spectrum $D_{\|,\|}(\vk) = D$.
We believe the above system is equivalent to \ens{5}{5.2}
in the strong coupling limit.
Note that \eno{akpz} is different from the KPZ equation\cite{kpz} by a
``missing''
term $(\grad_\perp u_\|)^2$. The exclusion of such a term is a
consequence of the {\it locality} of the dynamics in \vB.
This missing term gives rise to anisotropy
between the $\dagger$- and $\perp$- directions, which we explicitly
 take into account by allowing the diffusion coefficients
in the two directions to be different.

\eno{akpz} is straightforwardly analyzed by the DRG method.
We find the one-loop term in the recursion relation (\ref{rg}) to be
$I(d_z,d_\times) \sim d_z(19-3d_\times) - (16 - 9d_\times + d_\times^2)$. It
changes sign at $d_z^* \approx 0.24$ along the critical line $d=2$.
For $1\ge d_z > d_z^*$, we have $I(d=2) > 0$, and
the system flows to strong coupling as in the KPZ limit ($d_z=1$).
However, for $d_z<d^*_z$, the nonlinearity is irrelevant.
There, the series of approximations leading to
\eno{akpz} are no longer valid as $\vu_\perp$ and $u_\|$ become
comparable.  In fact, in the limit  $d_z=0$,
\eno{akpz} becomes the DDS with anisotropic noise\cite{js}, which actually
belongs to an universality class different from the $d_z=0$ limit of
\eno{5} (DDS with isotropic noise0).
Nevertheless, we expect the result for $d_z > d_z^*$ to be
good, and find that the KPZ-like strong coupling behavior
 dominating for $3/4$ of the way along the critical line.

To obtain the boundary of the strong coupling region away from the
critical line $d=2$, we will need to carry out the DRG calculation to
higher order.  However, we may use
the full one-loop result to get a ``feel" of the boundary in the
vicinity of the critical line.  Solving
for the root of $I(d_z^*,d_\times^*)=0$, we obtain a tentative
boundary which is sketched in
the space of generalized dimensions $(d_\times,d_z)$ to guide the eye
(see Figure 2). The result suggest that KPZ-like behavior is likely
to dominate a large portion of the $(d_\times,d_z)$-space.
It is then reasonable to expect the existence of
a phase transition to strong coupling in the physical dimension
$(d_\times, d_z) = (2,1)$.  However, numerical simulations of
\eno{akpz} in the physical dimension
 are needed to make a definitive conclusion.

In the remaining part of this paper, we shall assume that a phase
transition for the driven line liquid does exist in 3-dimensions
and explore its consequences. Pictorially, the transition is between
a line configuration described by the 3-dimensional generalization
of Figure 1a at small driving force
$J_e$, to the one described by the generalization of Figure 1b
when $J_e$ is increased beyond a critical value  $J_e^c$.
(The actual value $J_e^c \sim
\nu^{3/2}/D^{1/2} \rho''(B_0)$ will depend on the specifics of
material and bias conditions.)
If we view the lines as the streamlines of a fluid flow,
then the two phases correspond to
``laminar" and ``turbulent" flows respectively.
The laminar phase is described by the linear theory\cite{mn}.
However, the turbulent phase is much more complicated.
{}From experience\cite{kpz,surface} with the KPZ equation in
3-dimensions
$(d_\times,d_z) = (1,2)$, we expect  both the renormalized diffusion
coefficient and noise amplitude to diverge algebraically
in the infra-red limit $k\to 0$.
These anomalies will lead to
a singular structure factor.
They will also give anomalous dynamics ---
the response of the system will spread faster
then the usual $t^{1/2}$ for diffusion. Critical properties
associated with the transition itself are also of interest.
The nature of this type of dynamic phase transition has
been discussed in detail elsewhere\cite{dpt}.
Here we merely point out that
associated with a diverging length scale at the critical point, we
expect singularities in global quantities such as the
renormalized $E_0$ in \eno{5}, i.e., $E_0({\rm sing.}) \sim (J_e -
J_e^c)^{2-\alpha}$.
Such singularities should be detectable from simple I-V
measurements.

As mentioned at the beginning of this paper, we expect the
qualitative features of the
driven dynamics discussed here to be applicable to a generic
line liquid. It will for example be interesting to re-examine
the proliferation of vortex loops in experiments of
driven superfluid Helium\cite{He} in light of the phase transition discussed
here.
In addition to line liquids, the dynamics (\eno{akpz} in particular)
explored here provides a convenient link between the KPZ and DDS
dynamics, which are two of the simplest generalization of the
diffusion equation, and
have appeared in a wide variety of nonequilibrium
 problems. This link may be exploited to obtain a perturbative
access to the strong coupling fixed point itself.

The author would like to thank D. S. Fisher, M. Kardar, D. R.
Nelson, and J. P. Sethna for useful discussions. This work was
supported by NSF through grant No. DMR-90-96267 and by IBM.

\end{narrowtext}

\figure{
(a) Flux lines confined to a thin superconductor film.
Electric current $\vje$ is applied normal to the film. Lorentz force
{\bf F} drives the lines to the right.
(b) A ``scrambled" line configuration from the
contour plot of a {\it rough} surface.}
\figure{
The space of generalized dimensions $(d_\times,d_z)$:  The
dashed line is the line of critical dimensions $d_\times + d_z =
2$. The dotted line is a tentative boundary separating the KPZ-like
and the DDS-like dynamics (see text). The physical dimension of
interest is marked by the asterisk.}
\vfill
\newpage
\vfill
\psfig{figure=/home/cmts/hwa/text/REVTEX2/ddl2.fig1}
\vfill
\newpage
\vfill
\psfig{figure=/home/cmts/hwa/text/REVTEX2/ddl2.fig2}
\vfill

\begin{references}
\bibitem{note0} Point disorders, which preserves
superconductivity at low temperatures by collectively pinning the
flux lines\cite{vg}, are irrelevant in the liquid phase\cite{nl} and will be
neglected in this study.
\bibitem{note1} For layered cuprate superconductors, the
resistivity also depends on the angle between the applied field
and the $c$-axis. We shall not include this complication here as
it will not change the qualitative result of our study.
\end{references}
\end{document}